\documentclass[preprint,aps,showpacs,prd,tightenlines]{revtex4}
\usepackage{graphicx}
\usepackage[latin1]{inputenc}
\newcommand{\be}{\begin{eqnarray}} 
\newcommand{\ee}{\end{eqnarray}}
\newcommand{\nn}{~\nonumber \\}
\newcommand{\p}{{\cal P}\exp}

\newcommand{\ssh}{\gamma\cdot}

\newcommand{\bmp}{\noindent\begin{minipage}{16cm}}
\newcommand{\emp}{\end{minipage}\vskip 7mm} 
\newcommand{\unsplit}{\check}
\newcommand{\system}{\hat}
\newcommand{\kernel}{\bar}
\newcommand{\sigmaf}{\sigma\hspace{-1mm}:\hspace{-1mm}F}

\begin{document}


\title{
Gluon propagation in space-time dependent fields
}

\author{Dennis D. Dietrich}
\affiliation{
The Niels Bohr Institute, Copenhagen, Denmark
}
\affiliation{
Laboratoire de Physique Th\'eorique, Universit\'e Paris XI, Orsay, France
}


\begin{abstract}

The propagator for gluons in a space-time dependent field is derived. This
is accomplished by solving the equation of motion for the gluonic Green's
functions. Subsequently a relationship between the quark and the gluon 
propagator is
presented. With its help characteristics of the bosonic propagator can be
obtained from those of the fermionic and vice versa. Finally, this relation
is discussed for the special case of ultrarelativistic collisions
in the semiclassical limit.

\end{abstract}



\maketitle


\section{Introduction}

In Refs. \cite{dean,field,dddhelsinki} the full retarded fermion propagators in the
presence of arbitrarily space-time-dependent classical fields were studied. 
Here the generalisation to the corresponding gluon propagator is presented. 
Correlators of this kind are, for example, needed in order to study the 
production of the respective particles by vacuum polarisation
\cite{dean,field,dddhelsinki,ddd,baltz,lm,qqpa,qcd} or calculate induced currents
and condensates. Such phenomena are of importance for the physics of the early universe \cite{cornwall} and ultrarelativistic heavy-ion
collisions together with the quark-gluon plasma (QGP) \cite{kinetic}.
The strong classical fields are the common feature in these systems, while 
the detailed characteristics of the field configurations differ. After
providing the general formalism, this paper focusses on particle propagation
in ultrarelativistic collisions.

A lot of effort is made to study the QGP's production and equilibration
\cite{qm} in nuclear collision experiments at the Relativistic
Heavy-Ion Collider (RHIC) and the Large Hadron Collider (LHC), currently
under construction. 
It is a widely used assumption that the initial state in heavy-ion collisions 
is dominated by gluons. Due to their large occupation number and as a first
approximation one can begin the system's description with a classical 
background field. Quantum fluctuations have to be investigated subsequently.
Those considerations have also lead to the McLerran-Venugopalan (MV) model 
\cite{jimwlk}. There, the projectiles are represented by classically 
interacting colour-charge distributions on the two branches of the
light-cone. One observes that quantum corrections are enhanced
by kinematical factors. They can be included into the distributions by means
of the JIMWLK equation \cite{jimwlk}. 

In any case, the high occupation-number bosonic fields are that strong that
multiple couplings to the classical field are not suppressed even for a small
coupling constant $g$. In the absence of other scales they have to be 
taken into account to all orders. The leading quantum processes
concern terms in the classical action of second order in the fields
(fermions, antifermions, and bosonic quanta).
From those terms, the inverse of the respective two-point Green's functions
can be read off. Their solutions give the propagators of the corresponding 
particles to all orders in the classical field. Depending on the boundary 
conditions that are imposed, the correlator will serve to obtain expectation 
values (in-in-formalism, e.g., retarded and advanced propagator) or 
probabilities (in-out-formalism, e.g., Feynman propagator). In the following, 
the first possibility will be pursued mostly.

The paper is organised as follows: Section II contains the derivation of
the equation of motion for the gluon Green's functions within the framework
of the background-field method of QCD.
Section III derives the retarded gluon-propagator as a solution of the
equation of motion in background-field Feynman-gauge.
Section IV gives the retarded Green's function solution for the
quadratic Dirac-operator.
In section V the interrelation between the retarded gluon propagator and the
retarded quadratic Dirac-propagator is worked out. Additionally the
connection to the retarded linear Dirac-propagator is shown. These
results are used in order to translate the known characteristics of
the two-point function for the first-order Dirac-equation into those of
the other two correlators. Section VI applies the findings of the previous
sections to ultrarelativistic
heavy-ion collisions. The cases of nucleus-nucleus and nucleus-nucleon
collisions are addressed as well as that of the propagation of gluons with a
large momentum scale. In the last situation all terms for the retarded
propagator are spelled out and the Feynman propagator is available, too.
Section VII summarises the results. 

Throughout the paper the metric tensor is given by: $g^{\mu\nu}={\rm
diag}(1,-1,-1,-1)$, angular momenta are measured in units of $\hbar$, and
velocities in fractions of the speed of light $c$. From hereon, the coupling
constant is included in the definition of the classical field: 
\mbox{$gA^\mu_{old}=A^\mu_{new}$}. {$\vec v$} represents the three-vector of 
the spatial components of any four-vector $v$. The convention for Fourier
transformations of one-point functions is:
\be
f(x)
=
\int
\frac{d^4k}{(2\pi)^4}
e^{-ik\cdot x}
f(k),
\ee

\noindent
that for two-point functions:
\be
f(x,y)
=
\int
\frac{d^4p}{(2\pi)^4}
\frac{d^4q}{(2\pi)^4}
e^{-iq\cdot x}
e^{+ip\cdot y}
f(q,p).
\ee


\section{The equation of motion for gluon Green's functions}

In order to derive the requested equation of motion for the gluonic Green's
functions, let us start with the
classical action for the bosonic sector of the system:
\be
S=\frac{1}{g^2}\int d^4x {\cal L}_G
\ee

\noindent
with the Lagrangean density for the gauge field
\be
{\cal L}_G=\frac{1}{4}F^a\hskip -1mm:\hskip -1mmF^a
\ee

\noindent
where the field tensor is defined as, 
\be
F_{\mu\nu}=i[D_\mu,D_\nu]
\label{fieldtensor}
\ee

\noindent
with the covariant derivatives:
\be
D_\mu
=
\partial_\mu-i\unsplit A_\mu.
\ee

\noindent
The field tensor is explicitely given by:
\be
F^a_{\mu\nu}[\unsplit A]
=
\partial_\mu\unsplit A^a_\nu
-
\partial_\nu\unsplit A^a_\mu
+
f^{abc}\unsplit A^b_\mu\unsplit A^c_\nu.
\ee

The splitting of the gauge field $\unsplit A_\mu$ into a classical $A_\mu$ and 
a quantum $Q_\mu$ field:
\be
\unsplit A_\mu=A_\mu+Q_\mu,
\ee

\noindent
allows to express the field tensor as a functional of two fields:
\be
F^a_{\mu\nu}[A+Q]
=
F^a_{\mu\nu}[A]
+
D_\mu^{ab}[A]Q_\nu^b
-
D_\nu^{ab}[A]Q_\mu^b
+
f^{abc}Q_\mu^bQ_\nu^c.
\ee

\noindent
\mbox{$D_\mu^{ab}[A]$} stands for the covariant derivative as a functional of
the classical field $A_\mu$ only.
The Lagrangean density now is composed of a sum of terms containing different 
powers of the quantum fields. The terms of second order
determines the equation of motion for the propagator of the quantum
fluctuations in the classical background. It is given by:
\be
{\cal L}_G^{Q^2}
=
\frac{1}{2}\left\{
(D_\mu^{ab}[A]Q_\nu^b)(D^{ac\nu}[A]Q^{c\mu})
-
(D_\mu^{ab}[A]Q_\nu^b)(D^{ac\mu}[A]Q^{c\nu})
-
f^{abc}F^a_{\mu\nu}Q^{b\mu}Q^{c\nu}
\right\}.
\ee

Additional terms of second order in the quantum fluctuations arise
from the gauge-fixing term. The gauge fixing term to be added to the 
Lagrangean density reads for the background-field Feynman-gauge:
\be
{\cal L}_{GF}
=
{\cal L}_{GF}^{Q^2}
=
-\frac{1}{2}(D_\mu^{ab}[A]Q^{b\mu})(D_\nu^{ac}[A]Q^{c\nu})
\ee

As the Lagrangean density is always integrated over in order to obtain the
action, the expressions for the second order terms can be
transformed by virtue of partial integrations. The representation as
Lagrangean density is kept for the sake of simplicity:
\be
{\cal L}_G^{Q^2}
=
\frac{1}{2}
Q^{b\mu}
\left\{
D_\lambda^{ba}[A]D^{ac\lambda}[A]g_{\mu\nu}
-
D_\mu^{ba}[A]D^{ac}_\nu[A]
\right\}
Q^{c\nu}
\ee
\be
{\cal L}_{GF}^{Q^2}
=
\frac{1}{2}Q^{b\mu}D_\mu^{ba}[A]D_\nu^{ac}[A]Q^{c\nu}.
\ee

Finally, the differential operator for the equation of motion for the gluonic 
Green's functions can be read off by omitting the quantum fields. Thus the
equation of motion reads:
\be
\Gamma^{-1}(x)\Gamma_{\mu\nu}(x,y)
=
\delta^{(4)}(x-y)g_{\mu\nu}
\ee

\noindent
with the differential operator \footnote{The covariant derivatives as well
as the operator itself are elements of $U(N)$.}:
\be
\Gamma^{-1}
=
D[A]\cdot D[A].
\ee

Further the contribution to the Lagrangean density originating from the
Faddeev-Popov ghosts in background-field Lorenz-gauge is given by:
\be
{\cal L}_{FP}
=
-
(D_\mu^{ab}[A]\chi^{b\dagger})
(D_\mu^{ac}[A+Q]\chi^c)
\ee

\noindent
Partial integration yields:
\be
{\cal L}_{FP}
=
\chi^{b\dagger}D_\mu^{ba}[A]D_\mu^{ac}[A+Q]\chi^c
\ee

\noindent
Therefore, the equation of motion for the ghost propagator in the classical 
field reads:
\be
\Gamma^{-1}(x)\Gamma(x,y)
=
\delta^{(4)}(x-y).
\ee

\noindent
Note the connection between the quantum gluon and the ghost propagator in
this gauge:
\be
\Gamma_{\mu\nu}
=
\Gamma g_{\mu\nu},
\ee

\noindent
whence they can be used as synonymous to each other for most practical
purposes.


\section{The retarded gluon propagator}

First look at the homogeneous equation for the gluon (ghost) propagator:
\be
\{D[A(x)]\cdot D[A(x)]\}\Gamma_H(x,y)=0
\label{homogeneous}
\ee

\noindent
Explicitly one has:
\be
\{
{\partial_0}^2
-2i
A_0\partial_0
-
\Delta
+2i
\vec A\cdot\vec\partial
-i
(\partial\cdot A)
-
A\cdot A
\}(x)
\Gamma_H(x,y)
=
0
\label{explicitly}
\ee

\noindent
where the time derivatives have been singled out, because the boundary 
conditions for the retarded propagator are given for equal time. The
differential operator is taken at the space-time point $x$, which is only 
denoted once for the sake of brevity. $\Delta$
stands for the Laplace operator. In the term \mbox{$(\partial\cdot A)$} the 
derivative does not act accross the rounded brackets. The previous equation 
can be reexpressed as
\be
\system\Gamma^{-1}(x)\system\Gamma_H(x,y)=0
\label{system}
\ee

\noindent
with
\be
\system\Gamma^{-1}
=
\sigma_0
\partial_0
+
\kernel\Gamma^{-1}
\ee

\noindent
where
\be
\kernel\Gamma^{-1}
=
-i
\sigma_0
A_0
-
\sigma_+
+
\sigma_-
\{
-
\Delta
+2i
\vec A\cdot\vec\partial
-i
(\partial\cdot A)
-
A\cdot A
\}
+i
\sigma_3
A_0
\ee

\noindent
while
\be
2\system\Gamma_H
=
\sigma_0
(\Gamma_H^{(1)}+\partial_0\Gamma_H^{(2)})
+
2
\sigma_+
\Gamma_H^{(2)}
+
2
\sigma_-
\partial_0\Gamma_H^{(1)}
+
\sigma_3
(\Gamma_H^{(1)}-\partial_0\Gamma_H^{(2)})
\ee

\noindent
where $\Gamma_H^{(1)}$ and $\Gamma_H^{(2)}$ are two independent solutions
of Eq. (\ref{homogeneous}). $\sigma_j$
with \mbox{$j\in\{1,2,3\}$} are the Pauli matrices and $\sigma_0$ the
corresponding unit matrix. $\sigma_\pm$ are defined as
\mbox{$2\sigma_\pm=\sigma_1\pm i\sigma_2$}. Eq. (\ref{system}) is solved by:
\be
\system\Gamma_H(x,y)
=
\p
\left\{
\int_{y_0}^{x_0}d\xi_0 \kernel\Gamma^{-1}(\xi_0,\vec x)
\right\}.
\label{homogeneoussolution}
\ee

\noindent
Choosing a different boundary condition than 
\mbox{$\system\Gamma_H(x,y)=\sigma_0$} at \mbox{$x_0=y_0$} would only lead to 
overall factors to the right of the path-ordered exponential, which, for a
homogeneous differential equation does not lead to independent solutions.

The retarded gluonic propagator is to vanish for negative time-differences:
\mbox{$\Gamma(x,y)=0$} for \mbox{$x_0-y_0<0$}. Further, it is to be continuous at
\mbox{$x_0=y_0$} and there it is to have a discontinuous first temporal 
derivative: 
\mbox{$
\lim_{x_0\rightarrow y_0+0}\partial_0(x)\Gamma(x,y)
-
\lim_{x_0\rightarrow y_0-0}\partial_0(x)\Gamma(x,y)
=
\delta^{(3)}(\vec x-\vec y)
$}.  
Hence, the retarded propagator can be expressed as:
\be
\Gamma_R(x,y)
=
\system\Gamma_H^{(2)}(x,y)\theta(x_0-y_0)\delta^{(3)}(\vec x-\vec y)
\label{retardedgluonpropagator}
\ee

\noindent
Like in \cite{dean,field} this solution is fit to be analysed in various 
configurations with the help of the general resummation formula 
(\ref{genresum}) and
the group property (\ref{group}). However, here information on the
characteristics of the gluonic two-point function are to be obtained by
comparing it to quark correlators.


\section{The retarded quadratic fermion-propagator}

Starting out from the linear Dirac-equation
\be
\{i\gamma\cdot D[a(x)]-m\}G(x,y)
=
\delta^{(4)}(x-y)
\label{lineardirac}
\ee

\noindent
in the presence of the gauge field $a$ in fundamental representation, the 
quadratic differential-equation is obtained through the substitution
\be
G(x,y)
=
\{-i\gamma\cdot D[a(x)]-m\}
g(x,y)
\label{substitution}
\ee

\noindent
and one finds
\be
\{
D[a]\cdot D[a]
%
-\sigmaf
/2
+
m^2
\}(x)
g(x,y)
=
\delta^{(4)}(x-y)
\label{quadratic}
\ee

\noindent
where use has been made of the definition of the field tensor 
(\ref{fieldtensor}) and with the commutator of the $\gamma$ matrices 
\mbox{$2\sigma^{\mu\nu}=-i[\gamma^\mu,\gamma^\nu]$}.

In analogy to the gluonic case, the homogeneous part of this equation is 
equivalent to:
\be
\system g^{-1}(x)\system g_H(x,y)=0,
\ee

\noindent
with
\be
\system g^{-1}
=
\sigma_0\partial_0
+
\kernel g^{-1}
\label{quadraticsystem}
\ee

\noindent
where
\be
\kernel g^{-1}
=
-i
\sigma_0
a_0
-
\sigma_+
+
\sigma_-
\{
-
\Delta
+2i
\vec a\cdot\vec\partial
-i
(\partial\cdot a)
-
a\cdot a
-
\sigmaf[a]/2
+
m^2
\}
+i
\sigma_3
a_0
\ee

\noindent
while
\be
2\system g_H
=
\sigma_0
(g_H^{(1)}+\partial_0g_H^{(2)})
+
2
\sigma_+
g_H^{(2)}
+
2
\sigma_-
\partial_0g_H^{(1)}
+
\sigma_3
(g_H^{(1)}-\partial_0 g_H^{(2)})
\ee

\noindent
where $g_H^{(1)}$ and $g_H^{(2)}$ are two independent homogeneous solutions
of Eq. (\ref{quadratic}). Then Eq. (\ref{quadraticsystem}) is solved by:
\be
\system g_H(x,y)
=
\p\left\{
\int_{y_0}^{x_0}d\xi_0\kernel g^{-1}(\xi_0,\vec x)
\right\}
\ee

The boundary conditions in the fermionic case are
the same as in the bosonic, i.e., \mbox{$g(x,y)=0$} for \mbox{$x_0-y_0<0$},
\mbox{$\lim_{x_0\rightarrow y_0+0}g(x,y)=\lim_{x_0\rightarrow y_0-0}g(x,y)$},
and 
\mbox{$
\lim_{x_0\rightarrow y_0+0}\partial_0(x)g(x,y)
-
\lim_{x_0\rightarrow y_0-0}\partial_0(x)g(x,y)
=
\delta^{(3)}(\vec x-\vec y)
$}, whence the retarded propagator can be expressed as:
\be
\system g_R(x,y)
=
\system g_H^{(2)}(x,y)\theta(x_0-y_0)\delta^{(3)}(\vec x-\vec y).
\label{retardedquadraticfermionpropagator}
\ee


\section{Interrelation between the propagators}

With the help of the general resummation formula derived in \cite{field}:
\bmp
\be
&&\p\left\{\int^{x_0}_{y_0}d\xi_0[B(\xi_0)+C(\xi_0)]\right\}
=
\nn
&=&
\p\left\{\int^{x_0}_{y_0}d\xi_0 B(\xi_0)\right\}
\times
\nn
&&\times
\p\left[
\int^{x_0}_{y_0}d\xi_0
\p\left\{\int_{\xi_0}^{y_0}dz_0 B(z_0)\right\}
C(\xi_0)
\p\left\{\int^{\xi_0}_{y_0}dz_0 B(z_0)\right\}
\right]
=
\nn
&=&
\p\left[
\int^{x_0}_{y_0}d\xi_0
\p\left\{\int_{\xi_0}^{x_0}dz_0 B(z_0)\right\}
C(\xi_0)
\p\left\{\int^{\xi_0}_{x_0}dz_0 B(z_0)\right\}
\right]
\times
\nn
&&\times
\p\left\{\int^{x_0}_{y_0}d\xi_0 B(\xi_0)\right\},
\label{genresum}
\ee
\emp

\noindent
which is based on the group property valid for path-ordered exponentials:
\be
\p\left\{\int_{y_0}^{x_0}d\xi_0B(\xi_0)\right\}
=
\p\left\{\int_{z_0}^{x_0}d\xi_0B(\xi_0)\right\}
\times
\p\left\{\int_{y_0}^{z_0}d\xi_0B(\xi_0)\right\},
\label{group}
\ee

\noindent
the bosonic propagator can be identified in the fermionic. This is achieved
by making the choice \mbox{$C=c=\sigma_-\{-\sigmaf[a]/2+m^2\}$}. The
homogeneous solution of the quadratic Dirac-equation can then be reexpressed
as:
\be
\system g_H(x_0,y_0;\vec x)
=
\p\left\{+\int^{x_0}_{y_0}d\xi_0
\system\Gamma_H^\prime(x_0,\xi_0;\vec x)
c(\xi_0,\vec x)
\system\Gamma_H^\prime(\xi_0,x_0;\vec x)
\right\}
\system\Gamma_H^\prime(x_0,y_0;\vec x).
\ee

\noindent
The $\vec y$ in all the arguments has been left out, because due to the
$\delta^{(3)}(\vec x-\vec y)$ present in the propagator, it is always taken
equal to $\vec x$ in the end. The bosonic function has to be taken in the
fundamental representation. This supersymmetric replacement has been 
indicated by the prime.

Ultimately, the inverse relation is required, whence the task of identifying
the fermionic propagator in the bosonic arises. To this end, start out with
the bosonic homogeneous solution (\ref{homogeneoussolution}). First multiply with unity in the space
of the $\gamma$ matrices everywhere. The inverse operation is taking one
quarter of the trace over the elements of the Clifford algebra, which has not 
been denoted seperately for the sake of brevity. Afterwards, the application 
of the resummation formula (\ref{genresum}) with \mbox{$C=-c$} results in:
\be
\system\Gamma_H(x_0,y_0;\vec x)
=
\p\left\{-\int_{y_0}^{x_0}d\xi_0
\system g_H^\prime(x_0,\xi_0;\vec x)
c^\prime(\xi_0,\vec x)
\system g_H^\prime(\xi_0,x_0;\vec x)
\right\}
\system g_H^\prime(x_0,y_0;\vec x)
\ee

\noindent
The primes indicate that the fermionic functions are constructed in the
adjoint representation. The inserted matrix $c^\prime$ can be 
split into the contributions from the mass and the field tensor, respectively.
Resumming all terms involving the mass corresponds to replacing every massive
quark function $\system g^\prime_H$ by a massless one $\system g^\prime_h$:
\be
\system\Gamma_H(x_0,y_0;\vec x)
=
\p\left\{\frac{1}{2}\int_{y_0}^{x_0}d\xi_0
\system g_h^\prime(x_0,\xi_0;\vec x)
\sigmaf[A(\xi_0,\vec x)]\sigma_-
\system g_h^\prime(\xi_0,x_0;\vec x)
\right\}
\system g_h^\prime(x_0,y_0;\vec x)
\label{masslessconnection}
\ee

\noindent
Again this expression is fit to be studied with the help of Eqs.
(\ref{genresum}) and (\ref{group}) along the lines of Refs.
\cite{dean,field}. Here we shall concentrate on an expansion in the
insertions of the field tensor. To lowest order this yields for the relevant 
$\sigma_+$ component:
\be
\system\Gamma_H^{[0](2)}(x_0,y_0;\vec x)
=
\system g_h^{\prime(2)}(x_0,y_0;\vec x).
\ee

\noindent
The corrections arising in higher orders of the expansion include further
couplings to the field tensor. The $\sigma_+$ component of the first order 
reads:
\be
\system\Gamma_H^{[1](2)}(x_0,y_0;\vec x)
=
\frac{1}{2}\int_{y_0}^{x_0} d\xi_0 
\system g_h^{\prime(2)}(x_0,\xi_0;\vec x)
\sigmaf[A(\xi_0,\vec x)]
\system g_h^{\prime(2)}(\xi_0,y_0;\vec x),
\ee

\noindent
where use has been made of the relations 
\mbox{$\sigma_\pm\sigma_\mp\sigma_\pm=\sigma_\pm$} and 
\mbox{$(\sigma_\pm)^2=0$}. In fact, to all orders only the $\sigma_+$
component of $\system g_h^\prime$ contributes to $\system\Gamma_H^{(2)}$.
Multiplying the previous two equations with
\mbox{$\theta(x_0-y_0)\delta^{(3)}(\vec x-\vec y)$} in order to obtain the
contributions to the retarded propagator (\ref{retardedgluonpropagator}) 
leads to:
\be
\Gamma_R^{[0]}(x,y)
=
g_r^\prime(x,y)
\label{orderzero}
\ee

\noindent
and
\be
\Gamma_R^{[1]}(x,y)
=
\frac{1}{2}\int d^4\xi 
g_r^\prime(x,\xi)
\sigmaf[A(\xi)]
g_r^\prime(\xi,y),
\label{orderone}
\ee

\noindent
where \mbox{$\theta(x_0-y_0)=\theta(x_0-\xi_0)\theta(\xi_0-x_0)$} for
\mbox{$\xi_0\in[x_0,y_0]$} and afterwards the definition of the massless 
retarded quadratic fermion-propagator $g_r$ in accordance with Eq.
(\ref{retardedquadraticfermionpropagator}) have been exploited.
In vacuum configurations of the gauge field Eq. (\ref{orderzero})
becomes exact and the two propagators are identical.  
Last but not least, 
the retarded quadratic fermion-propagator $g_R$ can be constructed from the
retarded propagator $G_R$ for the linear Dirac-equation. Here the massless
case is needed:
\be
g_r(x,y)
=
-
\int d^4z
G_r(x,z)
G_r(z,y).
\label{convolution}
\ee

\noindent
which provides the link to the results presented in Refs.
\cite{dean,dddhelsinki,field}.
The minus sign originates from the sign of the covariant derivative in Eq. 
(\ref{substitution}).


\section{Semiclassical ultrarelativistic heavy-ion collisions}

The semiclassical description of an ultrarelativistic heavy-ion collision is
usually begun with a current of colour charges moving along the light-cone 
\cite{kr}. In order
to obtain the classical gauge field the Yang-Mills equations have to be
solved in the presence of this current. In Lorenz gauge the field 
consists of two different contributions. 
One of them is the radiation field inside the forward light-cone.
The other is made up of the Weizs\"acker-Williams (WW) sheets. These correspond
to the Coulomb fields of the different colour charges boosted into the
closure of the Lorentz group.
Therewith they become $\delta$-distributions in the light-cone coordinates.
This approximation is justified for the description of particles with low
longitudinal momentum (mid-rapidity). On the one hand they have a low resolution in this
direction. On the other they cannot be comovers of the charges on the
light-cone. For comoving particles the plane-wave expansion would not be
applicable \cite{baltz}.

In the Lorenz gauge for the classical field, the WW contribution to the 
gauge-field takes the form:

\bmp
\be
A_+^{WW}(x)
&=&
-\frac{g}{2\pi}
\sum_{n_L=1}^{N_L}
t_a(t_a^L)_{n_L}
\delta\left[x_--(b_-^L)_{n_L}\right]
\ln\lambda\left|\vec x_T-(\vec b_T^L)_{n_L}\right|,
\nn
A_-^{WW}(x)
&=&
-\frac{g}{2\pi}
\sum_{n_R=1}^{N_R}
t_a(t_a^R)_{n_R}
\delta\left[x_+-(b_+^R)_{n_R}\right]
\ln{\lambda\left|\vec x_T-(\vec b_T^R)_{n_R}\right|}.
\nn
A_T^{WW}(x)&=&0
\label{wwsheets}
\ee
\emp

\noindent
$\lambda$ is an
arbitrary constant, regularising the logarithm, which does not appear in 
quantities like the field tensor. The $t_a$ are the generators of $SU(3)_c$. 
The \mbox{$(t_a^{L,R})_n$} represent the colour of the charges.

In general, the presence of a second nucleus leads to a precession of the 
charges of the first nucleus and vice versa which is a manifestation of the
covariant conservation of the current. This causes further 
sheet-like contributions to the gauge field which can 
be combined with the WW fields by modifying the charges. 
Up to the next-to-leading order in perturbation theory they read \cite{kr}:

\bmp
\be
(t_a^L)_{n_L}
&=&
(t_a^L)_{n_L}^{in}
-
\alpha_S
\sum_{n_R=1}^{N_R}
f_{abc}(t_b^L)_{n_L}^{in}(t_c^R)_{n_R}^{in}
\theta\left[x_+-(b_+^R)_{n_R}\right]
\ln\lambda|\vec x_T-(\vec b_T^R)_{n_R}|
+
{\cal O}({\alpha_S}^2),
\nn
(t_a^R)_{n_R}
&=&
(t_a^R)_{n_R}^{in}
+
\alpha_S
\sum_{n_L=1}^{N_L}
f_{abc}(t_b^L)_{n_L}^{in}(t_c^R)_{n_R}^{in}
\theta\left[x_--(b_-^L)_{n_L}\right]
\ln\lambda|\vec x_T-(\vec b_T^L)_{n_L}|
+
{\cal O}({\alpha_S}^2).
\nn
\ee
\emp

\noindent
with the initial colours \mbox{$(t_a^{L,R})^{in}_{n_{L,R}}$}.
To all orders, the modifications amount to Wilson lines over the gauge field
along the branches of the light cone. Hence, the precession terms are absent
in adequate, i.e., light-cone gauges.
Colour neutrality requires
\mbox{$
\sum_{n_L=1}^{N_L}
t_a(t_a^L)_{n_L}^{in}
=
0
=
\sum_{n_R=1}^{N_R}
t_a(t_a^R)_{n_R}^{in}
$}.
Higher-order perturbative calculations for the gauge field as
solution of the Yang-Mills equations are only of value in the presence of
hard energy- or short time-scales. Up to now, non-perturbative solutions for
this problem have been obtained only in transverse lattice calculations
\cite{transverse}. In any case, the general form of the field---continuous
inside the forward light-cone and singular but integrable on the forward
light-cone---allows to express the retarded propagator by a finite
number of addends.

The final result for the massless retarded fermion propagator $G_r$ for the 
linear Dirac-equation in the ultrarelativistic limit\footnote{In the
ultrarelativistic limit, the "longitudinal impact parameters"
$(b_\pm^{R,L})_n$ are taken to be zero because of the longitudinal
contraction with the Lorentz factor $\gamma^{-1}\rightarrow 0$. While the 
correct limit of the propagator is thus obtained, subsequent applications of 
the result might be lacking terms because the order of the limits cannot be
exchanged (see, for example, the discussion in \cite{dean}).} 
reads \cite{dean}:
\be
iG_r(x,y)\gamma^0
=
+
G_h(x,y)
\delta^{(3)}(\vec x-\vec y)
\theta(x_0-y_0),
\label{masslessretardedlinear}
\ee

\noindent
with the homogeneous solution of the Dirac-equation:
\be
&&G_h(x,y)
=
G_h^{rad}(x,y)
+
\nn
&&+
\int d^4\xi
G_h^{rad}(x,\xi)
[T^L(\xi)\delta(\xi_-)+T^R(\xi)\delta(\xi_+)]
G_h^{rad}(\xi,y)
+
\nn
&&+
\int d^4\xi d^4\eta
G_h^{rad}(x,\xi)
T^L(\xi)\delta(\xi_-)
G_h^{rad}(\xi,\eta)
T^R(\eta)\delta(\eta_+)
G_h^{rad}(\eta,y)
+
\nn
&&+
\int d^4\xi d^4\eta
G_h^{rad}(x,\xi)
T^R(\xi)\delta(\xi_+)
G_h^{rad}(\xi,\eta)
T^L(\eta)\delta(\eta_-)
G_h^{rad}(\eta,y).
\label{urhichomogeneous}
\ee

\noindent
where for an even number of scattering centers in each projectile the
scattering matrices are given by
\be
T^L(x)+\rho^+
&=&
\rho^+
{\cal P}\prod_{n_L=1}^{N_L}\exp[-i\alpha_{n_L}^L(\vec x_T)],
\nn
T^R(x)+\rho^-
&=&
\rho^-
{\cal P}\prod_{n_R=1}^{N_R}\exp[-i\alpha_{n_R}^R(\vec x_T)],
\label{nuclearscatteringoperator}
\ee

\noindent
with
\be
2\pi\alpha_{n_{L,R}}^{L,R}=-gt_a(t_a^{L,R})_{n_{L,R}}
\ln\lambda\left|\vec x_T-(\vec b_T^{L,R})_{n_{L,R}}\right|,
\ee

\noindent
while the homogeneous solution in the radiation field is given by
\be
G^{rad}_h(x,y)
= 
\p\left\{
i\int_{y_0}^{x_0}d\xi_0 \gamma^0
[
i\gamma^j\partial_j(x)
+
\ssh A^{rad}(\xi_0,\vec x)]
\right\}.
\label{radiation}
\ee

\noindent
Higher orders in the $T$ cannot contribute to Eq. (\ref{urhichomogeneous}), 
because lines of constant $x_-$ or $x_+$ can only be crossed once 
\cite{dean,baltz}. 

This remains also true for the convolution of two retarded propagators in
Eq. (\ref{convolution}). There are at most contributions with one coupling 
to each scattering matrix in \mbox{$g_r(x,y)$}.

~\\

As already mentioned above, in nucleus-nucleus collsions the radiation field 
$A^{rad}$ is only known numerically up to now \cite{transverse}. This problem 
can be circumvented in proton-nucleus collisions by regarding the charge 
density of the proton as small
\cite{pa}. The Yang-Mills equations can be solved to all orders in the
charge density of the nucleus and to the first order in the charge density
of the proton.\footnote{In \cite{balitsky} the "expansion parameter" is not
the proton's charge density but the order of hatched commutators between the
WW fields of the two projectiles.} The propagator in the 
radiation field is replaced by its
perturbative expansion up to the first order in the thus obtained radiation
field. The scattering matrices for said case are obtained from the above 
expressions by keeping the nuclear scattering matrix \mbox{$T^L(x)$} as 
it is and by replacing the one for the proton by the lowest non-trivial order:
\be
T^R_{[1]}(x)
=
-i\rho^-\sum_{n_R=1}^{N_R}\alpha_{n_R}^R(\vec x_T).
\ee

\noindent
Note that the addend where the radiation field and the protonic scattering
matrix contribute simultaneously, is discarded, because it is of
second order in the charge density of the nucleon. 

~\\

In the case where a large momentum scale is assigned to the propagated
particle, be it in nucleus-nucleus or in proton-nucleus collisions, the
equations simplify further \cite{kr,lm}. Also the other scattering matrix
in Eq. (\ref{nuclearscatteringoperator}) is reduced to its lowest order:
\be
T^L_{[1]}(x)
=
-i
\rho^+
\sum_{n_L=1}^{N_L}\alpha_{n_L}^L(\vec x_T).
\ee

\noindent
The Yang-Mills equations are only solved perturbatively to the first order in 
each of the charge densities to yield the radiation field $A^{rad}_{[1]}$.
The propagator in the radiation field is expanded to the first order in the
field. At the end, only couplings to
the sheet-like contributions {\it or} to the radiation field are kept. Taking them
into account simultaneously, leads to higher order terms. Thus the retarded
linear Dirac-propagator to this order is given by Eq. 
(\ref{masslessretardedlinear}) with:
\be
&&G_h^{\rm II}(x,y)
=
\underline{G_h^0(x-y)}
+
\nn
&&+
\int d^4\xi
G_h^0(x-\xi)
[
\underline{T^L_{[1]}(\xi)\delta(\xi_-)}
+
\underline{T^R_{[1]}(\xi)\delta(\xi_+)}
-
i\gamma^0\ssh a^{rad}_{[1]}(\xi)]
G_h^0(\xi-y)
+
\nn
&&+
\int d^4\xi d^4\eta
G_h^0(x-\xi)
T^L_{[1]}(\xi)\delta(\xi_-)
G_h^0(\xi-\eta)
T^R_{[1]}(\eta)\delta(\eta_+)
G_h^0(\eta-y)
+
\nn
&&+
\int d^4\xi d^4\eta
G_h^0(x-\xi)
T^R_{[1]}(\xi)\delta(\xi_+)
G_h^0(\xi-\eta)
T^L_{[1]}(\eta)\delta(\eta_-)
G_h^0(\eta-y).
\label{homogeneouslargemass}
\ee

\noindent
\mbox{$G_h^0(x-y)$} stands for the free homogeneous solution of the linear
Dirac-equation without mass. Putting this into Eq. (\ref{convolution}) and subsequently
into Eq. (\ref{orderzero}) yields:

\bmp
\be
g_r^{\rm II}(x,y)
&=&
-g_r^0(x-y)
-\int d^4z G_r^0(x-z)G_r^{\rm II}(z,y)
-\int d^4z G_r^{\rm II}(x,z)G_r^0(z-y)
+
\nn
&&+
\int d^4\xi d^4\eta G_r^0(x-\xi)
T^L_{[1]}(\xi)\delta(\xi_-)
g_r^0(\xi-\eta)
T^R_{[1]}(\eta)\delta(\eta_+)
G_r^0(\eta-y)
+
\nn
&&+
\int d^4\xi d^4\eta G_r^0(x-\xi)
T^R_{[1]}(\xi)\delta(\xi_-)
g_r^0(\xi-\eta)
T^L_{[1]}(\eta)\delta(\eta_+)
G_r^0(\eta-y)
\ee
\emp

\noindent
with
\be
g_r^0(x,y)
=
-
\int
d^4z
G_r^0(x-z)
G_r^0(z-y)
\ee

For the use with Eq.
(\ref{orderone}) the addends which are not underlined will not contribute at
all. In connection with the radiated field tensor merely the free 
propagator in Eq. (\ref{homogeneouslargemass}) is needed. Hence one finds, 
still with some higher order terms to be removed in the first addend:
\be
\Gamma_R^{[1]}(x,y)
&=&
\frac{1}{2}\int d^4\xi
g_r^0(x-\xi)\sigmaf[A^{WW}(\xi)]g_r^0(\xi-y)
+
\nn
&&+
\frac{1}{2}\int d^4\xi
\underline{g_r^\prime}(x,\xi)\sigmaf[A^{WW}(\xi)]g_r^0(\xi-y)
+
\nn
&&+
\frac{1}{2}\int d^4\xi
g_r^0(x-\xi)\sigmaf[A^{WW}(\xi)]\underline{g_r^\prime}(\xi,y)
+
\nn
&&+
\frac{1}{2}\int d^4\xi
g_r^0(x-\xi)\sigmaf[A^{rad}_{[1]}(\xi)]g_r^0(\xi-y)
\ee 

\noindent
with:
\be
\underline{g_r}(x,y)
&=&
+
\int d^4\xi g_r^0(x-\xi)
[{T^L_{[1]}(\xi)\delta(\xi_-)}+{T^R_{[1]}(\xi)\delta(\xi_+)}]
G_r^0(\xi-y)
+
\nn
&&+
\int d^4\xi G_r^0(x-\xi)
[{T^L_{[1]}(\xi)\delta(\xi_-)}+{T^R_{[1]}(\xi)\delta(\xi_+)}]
g_r^0(\xi-y).
\ee

To this accuracy, the last contribution comes from the second
order term of Eq. (\ref{masslessconnection}):
\be
\Gamma_R^{[2]}(x,y)
&=&
\frac{1}{4}\int d^4\xi d^4\eta 
g_r^0(x-\xi)
\sigmaf[A_+^{WW}(\xi)]
g_r^0(\xi-\eta)
\sigmaf[A_-^{WW}(\eta)]
g_r^0(\eta-y)
+
\nn
&&+
\frac{1}{4}\int d^4\xi d^4\eta 
g_r^0(x-\xi)
\sigmaf[A_-^{WW}(\xi)]
g_r^0(\xi-\eta)
\sigmaf[A_+^{WW}(\eta)]
g_r^0(\eta-y),
\label{ordertwo}
\ee

\noindent
where only the field tensors as functionals of the sheet-like
configurations (\ref{wwsheets})\cite{kr}
\be
F_{+T}[A^{WW}_+]
=
\frac{g}{2\pi}
\sum_{n_L=1}^{N_L}
t_a(t_a^L)_{n_L}
\delta\left[x_--(b_-^L)_{n_L}\right]
\frac
{\vec x_T-(\vec b_T^L)_{n_L}}
{\left|\vec x_T-(\vec b_T^L)_{n_L}\right|^2}
\ee 

\noindent
and 
\be
F_{-T}[A^{WW}_-]
=
\frac{g}{2\pi}
\sum_{n_R=1}^{N_R}
t_a(t_a^R)_{n_R}
\delta\left[x_+-(b_+^R)_{n_R}\right]
\frac
{\vec x_T-(\vec b_T^R)_{n_R}}
{\left|\vec x_T-(\vec b_T^R)_{n_R}\right|^2}
\ee

\noindent
and all other components equal to zero, contribute.

Here, as at most two field insertions of any kind per addend are allowed, the
corresponding Feynman propagator can be obtained by replacing the free
retarded propagators by free Feynman propagators.

~\\

Gauge independent results can be obtained from an operator in different
ways, for example, by multiplying with adequate gauge links or by averaging 
over all gauges \cite{dean}. The MV-like models fall back on averaging over an
ensemble of charge distributions with a gauge-invariant weight
\cite{jimwlk}. This averaging can be carried out equivalently over the WW
fields \cite{km}.

In the case of a gluon propagating at a large momentum scale, the radiated
field is a functional linear in $A_+^{WW}$ and $A_-^{WW}$. Thus the same
holds for the gluon propagator in this situation. Due to the colour
neutrality of each projectile, the average over the charge densities or
equivalently the WW fields are to vanish \mbox{$\left<A_\pm^{WW}\right>=0$}.
Therefore, the average propagator to this order coincides with the free
propagator. The deviations from the free case start but at the next order or
if the average over an operator is taken that contains higer powers of the
propagator.

Along the same lines, the propagator in the proton-nucleus case is, per
definition of the approximation, a functional linear in the charge density
or the WW field of the proton. This does not change if terms beyond Eq.
(\ref{orderone}) are taken into account. Hence, after taking the average
over the proton imposing colour neutrality, only the average over the 
propagator in the field
of the nucleus is left. As before, deviations originating from the presence
of the nucleon only start playing a r\^ole in the following orders or in
averages over powers of the propagator.


\section{Summary}

Solutions for the retarded gluonic propagator in background-field Feynman
gauge and the retarded propagator for the quadratic Dirac-equation have been
derived to all orders in the classical field. Imposing the corresponding
boundary conditions has singled out the retarded correlators. Subsequently,
relations expressing each of these propagators as a functional of the other
have been determined. The one constructing the bosonic two-point function
from the fermionic has been appended with another one, linking the quadratic
fermion propagator with the linear one. With the help of said equations, the
results for Green's functions of the linear Dirac-operator from Refs.
\cite{dean,dddhelsinki,field} can be and have been transferred to those of 
the quadratic and the bosonic operator. 

This fact has been exploited for the case of ultrarelativistic collisions. For nucleus-nucleus collisions all
pieces of information, which do not require numerical calculations for the 
radiated field have been exposed. For nucleon-nucleus collisions, 
one can go further, because they allow for an expansion of the Yang-Mills
equations in the charge density of the proton. The building blocks for the 
description of this system have been provided. 

As a third system, the 
propagation of a gluon with a large momentum scale is investigated. In this 
case the Yang-Mills equations can be expanded in the densities of both
projectiles. All possible terms up to first order in any of the charge 
densities have been derived. In this case, also the Feynman propagator can
be obtained. At this order, averaging the propagators over the charge 
distributions along the line of the MV model, gives the free
propagator. Deviations arise when averaging over operators containing
powers of the propagator at this order or over the propagator at an higher
order. In the case of a proton-nucleus collision, the averaging procedure
leads to the average over the propagator in the field of the nucleus. 

It will be interesting to apply the presented correlators to describe
phenomena like vacuum polarisation and quantities like induced currents or
condensates in the gluonic case. Such investigations will be presented 
elsewhere.


\section*{Acknowledgements}

The authors stay at the LPT had been financed by the DAAD (German Academic 
Exchange Service).



\end{document}